\documentclass[twocolumn,showpacs,amsmath,amssymb,aps,pra]{revtex4}
\usepackage{graphicx}
\usepackage{bm}

\begin{document}

\title{Effective nonlinear Hamiltonians in dielectric media}

\author{J. A. Crosse}
\email{jac00@imperial.ac.uk}
\author{Stefan Scheel}
\email{s.scheel@imperial.ac.uk}
\affiliation{Quantum Optics and Laser Science, Blackett Laboratory, 
Imperial College London, Prince Consort Road, London SW7 2AZ}

\date{\today}

\begin{abstract}
We derive an effective Hamiltonian for the nonlinear process of
parametric down conversion in the presence of absorption. Based upon
the Green function method for quantizing the electromagnetic field, we
first set up Heisenberg's equations of motion for a single atom driven
by an external electric field and in the presence of an absorbing
dielectric material. The equations of motion are then solved to second
order in perturbation theory which, in rotating-wave approximation,
yields the standard effective interaction Hamiltonian known from
free-space nonlinear optics. In a second step, we derive the
local-field corrected Hamiltonian for an atom embedded in a dielectric
host medium, i.e. a nonlinear crystal. Here we show that the resulting
effective Hamiltonian is found to be trilinear in the electric and noise
polarization fields, and is thus capable of describing nonlinear noise
processes. Furthermore, it reduces to the phenomenological
nonlinear Hamiltonian for the cases where absorption, and hence 
the noise polarization field, vanishes.  
\end{abstract}


\maketitle
\section{Introduction}
\label{intro}

Ever since the discovery of second harmonic generation by Franken in
1961 \cite{franken}, nonlinear optical process have been the subject
of great interest. Uses of such processes cover the full spectrum of
possible applications from optical communications \cite{comm} at one
end to fundamental tests of quantum mechanics
\cite{bellini1,bellini2} at the other. The strongly correlated photons
that are created in these processes are regularly used in many quantum
cryptographic protocols \cite{pdcqi1,pdcqi2}, and in the areas of
quantum information processing and quantum computing \cite{qirev}. As
a result, these process have been the subject of much study
(for a necessarily incomplete selection, see
e.g. Refs. \cite{pdc1,pdc2,pdc3,pdc4}).

The fundamental theory that describes the interaction of light and
matter is quantum electrodynamics (QED). This theory has proven to be
highly successful over the past sixty years, and has accurately
described a disparate range of physical phenomena over a wide range of
energies, from scattering of charged particles in high energy colliders
to the low energy dynamics of atoms in electromagnetic fields. The
theory which, in its microscopic form, describes the interaction of
electromagnetic fields with charged particles, predicts the appearance
of nonlinear processes when high intensity fields interact with
certain types of matter. However, owing to the complexities associated
with the microscopic structure of matter, the calculation of such
properties are highly involved and often neglect some of the more
complicated features. For example, in such calculations absorption is
almost always neglected. As a result the standard approach to
nonlinear optical processes is, in the main, phenomenological and,
although it provides a good approximation at high intensities
\cite{bloem}, does not necessarily hold for all
situations. Furthermore, there are several circumstances where
nonlinear absorption is thought to be a critical factor
\cite{limit1,limit2} and hence the standard approach is insufficient
in these cases.

It is certainly the case that absorption plays an important role in
many physical processes. Recently, a method for consistently
quantizing the electromagnetic field in absorbing linear electric
and magnetic materials has been developed (for reviews, see e.g.
Refs.~\cite{linearreview,acta}).
There have been some attempts to extend this theory to
nonlinear materials \cite{schmidt1,nonlinear1,nonlinear2}, however, as
yet a full theory has proven to be elusive. 

In this article we present an extension of the linear quantum theory of
light in absorbing media to nonlinear processes. In the following we
will consider, as an example, the second order process of parametric
down conversion, where an input (pump) photon is converted by a
nonlinear medium to give two output photons (signal and idler) whose
frequencies sum to that of the input photon. We will begin, in
Sec.~\ref{sec:quant}, by briefly reviewing the quantization scheme for
linear absorbing dielectric materials. In Sec.~\ref{sec:eom}, we
consider the interaction of photons with a single atom to second order
in perturbation theory and derive an effective interaction Hamiltonian
for this process. In Sec.~\ref{sec:lfc}, in order to find the nonlinear
response for a bulk material,  we apply local field corrections to the
Green functions of the interacting electric fields by considering the
effect of placing the interacting atom in a cavity within a bulk
material. This method produces an effective interaction Hamiltonian for the
second order nonlinear process that includes the sought nonlinear
noise processes. Concluding remarks are given in
Sec.~\ref{sec:sum}. Some useful expressions and lengthy derivations
can be found in the Appendices.

\section{Electromagnetic field quantization in linear media} 
\label{sec:quant}

Before we outline the theory of electromagnetic field quantization
in nonlinear dielectric media we shall briefly review the quantization
of the electromagnetic field in a linearly responding medium
\cite{linearreview,acta}. We begin with the classical Maxwell
equations in frequency space. In the absence of free currents and
charges these equations take the form
\begin{gather}
\bm{\nabla}\cdot\mathbf{B}(\mathbf{r},\omega) = 0,\\
\bm{\nabla}\times\mathbf{E}(\mathbf{r},\omega) -
i\omega\mathbf{B}(\mathbf{r},\omega) = \mathbf{0},\\ 
\bm{\nabla}\cdot\mathbf{D}(\mathbf{r},\omega) = 0,\\
\bm{\nabla}\times\mathbf{B}(\mathbf{r},\omega) +
i\omega\mu_0\mathbf{D}(\mathbf{r},\omega) = \mathbf{0}, 
\end{gather}
with
\begin{equation}
\mathbf{D}(\mathbf{r},\omega) =
\varepsilon_0\mathbf{E}(\mathbf{r},\omega) +
\mathbf{P}(\mathbf{r},\omega). 
\end{equation}
Although magnetoelectric media can be treated along the same lines,
for simplicity we assume that there are no magnetic responses present.
For spatially local isotropic media, the general form of the linear
polarization field is
\begin{equation}
\label{eq:pol}
\mathbf{P}(\mathbf{r},\omega) =
\varepsilon_0\chi^{(1)}(\mathbf{r},\omega)
\mathbf{E}(\mathbf{r},\omega) +
\mathbf{P}_\mathrm{N}(\mathbf{r},\omega). 
\end{equation}
The first term in Eq.~(\ref{eq:pol}) is the linear response of the
medium to an external electric field with a linear susceptibility
$\chi^{(1)}(\mathbf{r},\omega)$. The second term is the linear
noise polarization field which describes Langevin noise that is
associated with absorption, and is required for the theory to be
consistent with the fluctuation-dissipation theorem. As a result the
frequency components of the electric field obey the inhomogeneous
Helmholtz equation
\begin{equation}
\bm{\nabla}\times\bm{\nabla}\times\mathbf{E}(\mathbf{r},\omega) -
\frac{\omega^2}{c^2}\varepsilon(\mathbf{r},\omega)
\mathbf{E}(\mathbf{r},\omega)
= \omega^2\mu_0\mathbf{P}_\mathrm{N}(\mathbf{r},\omega), 
\end{equation}
where $\varepsilon(\mathbf{r},\omega)=1+\chi^{(1)}(\mathbf{r},\omega)$
is the complex permittivity of the medium. This equation can be
formally solved using the Green tensor for the Helmholtz operator 
\begin{equation}
\mathbf{E}(\mathbf{r},\omega) = \omega^2\mu_0\int d^3s\,
\bm{G}(\mathbf{r},\mathbf{s},\omega)
\cdot\mathbf{P}_\mathrm{N}(\mathbf{s},\omega).
\end{equation}
The Green tensor $\bm{G}(\mathbf{r},\mathbf{s},\omega)$ solves the
Helmholtz equation with a point source
\begin{equation}
\bm{\nabla}\times\bm{\nabla}\times\bm{G}(\mathbf{r},\mathbf{s},\omega) -
\frac{\omega^2}{c^2}\varepsilon(\mathbf{r},\omega)
\bm{G}(\mathbf{r},\mathbf{s},\omega)
= \bm{\delta}(\mathbf{r}-\mathbf{s}).
\end{equation}

Quantization is then performed by relating the noise polarization field
to a set of bosonic field operators 
\begin{equation}
\label{eq:noisepol}
\hat{\mathbf{P}}_\mathrm{N}(\mathbf{r},\omega) =
i\sqrt{\frac{\hbar\varepsilon_0}{\pi}\varepsilon''(\mathbf{r},\omega)}\,
\hat{\mathbf{f}}(\mathbf{r},\omega) 
\end{equation}
and imposing canonical commutation relations for them,
\begin{equation}
\left[\hat{\mathbf{f}}(\mathbf{r},\omega),
\hat{\mathbf{f}}^{\dagger}(\mathbf{s},\omega')\right] =
\bm{\delta}(\mathbf{r}-\mathbf{s})\delta(\omega-\omega'). 
\end{equation}
Thus the frequency components of the quantized electric field can be
written as
\begin{equation}
\hat{\mathbf{E}}(\mathbf{r},\omega) =
i\sqrt{\frac{\hbar\varepsilon_0}{\pi}}\frac{\omega^2}{c^2\varepsilon_0}
\int d^3s \,\sqrt{\varepsilon''(\mathbf{s},\omega)}
\bm{G}(\mathbf{r},\mathbf{s},\omega)\cdot
\hat{\mathbf{f}}(\mathbf{s},\omega)
\end{equation}
and the total field operator reads
\begin{equation}
\hat{\mathbf{E}}(\mathbf{r}) = \int\limits^{\infty}_0 d\omega\,
\hat{\mathbf{E}}(\mathbf{r},\omega) + \mbox{h.c.}.
\end{equation}
The bosonic operators $\hat{\mathbf{f}}(\mathbf{s},\omega)$ and
$\hat{\mathbf{f}}^{\dagger}(\mathbf{s},\omega)$ describe collective
excitations of the electromagnetic field and the absorbing dielectric
material and can be viewed as the generalization of the free space
photonic mode operators to arbitrary media. The bilinear Hamiltonian  
\begin{equation}
\label{eq:HF}
\hat{H}_\mathrm{F} = \int d^3r\int\limits_0^\infty d\omega\,
\hbar\omega\, \hat{\mathbf{f}}^{\dagger}(\mathbf{r},\omega)\cdot
\hat{\mathbf{f}}(\mathbf{r},\omega),
\end{equation}
generates the time-dependent Maxwell equations from Heisenberg's
equations of motion for the electromagnetic field operators. 

This quantization scheme has already been successfully applied to a 
wide range of linear media including magnetic and magnetodielectric
materials and has been used to study many physical effects including 
spontaneous relaxation rates, atom surface interactions and various 
cavity QED processes \cite{acta}.

\section{Equations of motion of the light-atom system}
\label{sec:eom}

Starting with this theory, it is possible to study nonlinear
interactions of light with an atom in the presence of dielectric
bodies. Here, one solves the coupled equations of motion 
recursively to obtain an expansion in powers of the electric field
operator. Each of these higher-order terms corresponds to a specific
nonlinear process. From these interaction terms effective nonlinear
Hamiltonians can be derived which characterise each of these
processes. In this section we will look at the derivation of the
effective Hamiltonian for the second order nonlinear process of
parametric down conversion, where an input (pump) photon with
frequency $\omega_{p}$ is converted by a nonlinear crystal to give two
output photons (a signal photon with frequency $\omega_{s}$ and an
idler photon with frequency $\omega_{i}$) such that
$\omega_{p} = \omega_{s} + \omega_{i}$.

The interaction between light and a single atom can be described using
the multipolar coupling in the dipole approximation. Here the
electromagnetic field couples linearly to the dipole moment of the
atom. The multipolar coupling Hamiltonian can be written as 
\begin{equation}
\label{eq:linearH}
\hat{H} = \hat{H}_\mathrm{F} + \hat{H}_\mathrm{A} +
\hat{H}_\mathrm{int} 
\end{equation}
with
\begin{align}
\hat{H}_\mathrm{A} &= \sum_{i}\hbar\omega_{i}\hat{\sigma}_{ii}\,,
\nonumber\\ 
\hat{H}_\mathrm{int} &= -\hat{\mathbf{d}} \cdot
\hat{\mathbf{E}}(\mathbf{r}_A) 
\nonumber\\
&= -i\sum_{ij}\hat{\sigma}_{ij}d_{\mu ,ij}
\int d^{3}s\int d\omega'\sqrt{\frac{\hbar\varepsilon_0}{\pi}}
\frac{\omega'^{2}}{c^2\varepsilon_0}\nonumber\\
&\quad\times\sqrt{\varepsilon''(\mathbf{s},\omega ')}
G_{\mu\lambda}(\mathbf{r}_{A},\mathbf{s},\omega ')
\hat{f}_{\lambda}(\mathbf{s},\omega ') + \mbox{h.c.}
\end{align}
and $\hat{H}_\mathrm{F}$ from Eq.~(\ref{eq:HF}) above.
The above Hamiltonian has been written in component form 
with the Greek indices running over the three Cartesian coordinates. 
To these indices the summation convention applies. 
The atomic Hamiltonian is the sum of the projectors onto the
(undisturbed) energy levels of the free atom located at $\mathbf{r}_{A}$,
$\hat{\sigma}_{ii}=|i\rangle\langle i|$ is the projector onto the
$i$th eigenstate with energy $\hbar\omega_{i}$ and
$\hat{\sigma}_{ij}=|i\rangle\langle j|$ is the atomic flip operator
between the $i$th and $j$th atomic energy state. Note that we do not
apply the index summation convention over the (Latin) atomic state
indices.

From the Hamiltonian (\ref{eq:linearH}) we obtain Heisenberg's
equations of motion for the atomic and bosonic field operators as
\begin{align}
\label{eomrad}
&\dot{\hat{f}}^{\dagger}_{\lambda}(\mathbf{r},\omega) =
i\omega\hat{f}^{\dagger}_{\lambda}(\mathbf{r},\omega) - i\sum_{ij}
g_{\lambda ,ij}(\mathbf{r}_{A},\mathbf{r},\omega)\hat{\sigma}_{ij},\\
\label{eomatom}
&\dot{\hat{\sigma}}_{ij} =
i\omega_{ij}\hat{\sigma}_{ij} - i\sum_{k}\int d^{3}s
\int d\omega\nonumber\\
&\times\left\{\left[g_{\lambda ,jk}(\mathbf{r}_{A},\mathbf{s},\omega)
\hat{\sigma}_{ik} - g_{\lambda ,ki}(\mathbf{r}_{A},\mathbf{s},\omega)
\hat{\sigma}_{kj}\right]\hat{f}_{\lambda}(\mathbf{s},\omega)\right.\nonumber\\
&\left.\quad + \left[g^{\ast}_{\lambda ,jk}(\mathbf{r}_{A},\mathbf{s},\omega)
\hat{\sigma}_{ik} - g^{\ast}_{\lambda ,ki}(\mathbf{r}_{A},\mathbf{s},\omega)
\hat{\sigma}_{kj}\right]\hat{f}^{\dagger}_{\lambda}(\mathbf{s},\omega)\right\}
\end{align}
where $\omega_{ij} = \omega_{i} - \omega_{j}$ are the atomic
transition frequencies and the coupling constants
$g_{\lambda ,ij}(\mathbf{r},\mathbf{s},\omega)$
are defined by
\begin{equation}
g_{\lambda ,ij}(\mathbf{r},\mathbf{s},\omega) =
\frac{i}{\sqrt{\hbar\varepsilon_0\pi}}
\frac{\omega^2}{c^2}\sqrt{\varepsilon''(\mathbf{s},\omega)}
\,d_{\mu ,ij}G_{\mu\lambda}(\mathbf{r},\mathbf{s},\omega).
\end{equation}
The differential equations (\ref{eomrad}) and (\ref{eomatom})
completely describe the dynamics of the coupled light-atom system. As
we are primarily interested in the dynamics of the radiation field,
the next step will be to remove the atomic degrees of freedom. This is
done by formally solving for the atomic operators and resubstituting
the result into the equation for the field operators. The result will
be a single dynamic equation for the radiation field in the presence
of the atom. The formal solution of Eq.~(\ref{eomatom}) is
\begin{align}
&\hat{\sigma}_{ij}(t) = \hat{\sigma}_{ij}(0)e^{i\omega_{ij}t} - i\int\limits_0^{t}dt'\int d^{3}s\int d\omega\sum_{k} e^{i\omega_{ij}(t-t')}\nonumber\\ 
&\times\left\{\left[g_{\lambda ,jk}(\mathbf{r}_{A},\mathbf{s},\omega)\hat{\sigma}_{ik}(t') - g_{\lambda ,ki}(\mathbf{r}_{A},\mathbf{s},\omega)\hat{\sigma}_{kj}(t')\right]\hat{f}_{\lambda}(\mathbf{s},\omega)\right.\nonumber\\
&+\left.\left[g^{\ast}_{\lambda ,jk}(\mathbf{r}_{A},\mathbf{s},\omega)\hat{\sigma}_{ik}(t')-g^{\ast}_{\lambda ,ik}(\mathbf{r}_{A},\mathbf{s},\omega)\hat{\sigma}_{kj}(t')\right]\hat{f}^{\dagger}_{\lambda}(\mathbf{s},\omega)\right\}
\label{eq:sigma}
\end{align}
where $\hat{\sigma}_{ij}(0)$ is the atomic operator at $t=0$. This is
a recursive expression for $\hat{\sigma}_{ij}$. The full solution can 
be obtained by substituting Eq. (\ref{eq:sigma}) back into itself. The
result is an infinite series in increasing powers of the field
operators $\hat{f}_{\lambda}(\mathbf{s},\omega)$ and
$\hat{f}^{\dagger}_{\lambda}(\mathbf{s},\omega)$. Each of these higher
order terms corresponds to a specific nonlinear process that can
occur when the radiation field interacts with the atom. We are
interested in parametric down conversion and hence are interested in
the term that is quadratic in the field operators. In fact, since the 
parametric down conversion process creates two photons we will 
be interested in terms quadratic in field creation operators. It transpires 
that terms containing the annihilation operator will average to 
zero when we apply the rotating wave approximation. In the light of this, for notational clarity,
we will henceforth not display in detail terms that contain either 
$\hat{f}_{\lambda}(\mathbf{s},\omega)$ and/or $\hat{f}_{\mu}(\mathbf{s}',\omega')$.
Resubstituting Eq.~(\ref{eq:sigma}) twice and then inserting it back into
Eq.~(\ref{eomrad}) gives
\begin{widetext}
\begin{multline}
\dot{\hat{f}}^{\dagger}_{\nu}(\mathbf{r},\omega '') =
i\omega ''\hat{f}^{\dagger}_{\nu}(\mathbf{r},\omega '') - i\sum_{ijkp}
g_{\nu ,ij}(\mathbf{r}_{A},\mathbf{r},\omega)\bigg[\hat{\sigma}_{ij}(0)
e^{i\omega_{ij}t}
- i\int\limits_0^ tdt'\int\limits_0^{t'}dt''\int d^3s\int d^3s'
\int d\omega\int d\omega' e^{i\omega_{ij}(t-t')}
\\ 
\times\bigg\{g^{\ast}_{\lambda ,kj}(\mathbf{r}_{A},\mathbf{s},\omega) 
\left(\hat{\sigma}_{ik}(0)e^{i\omega_{ik}t'}
-ie^{i\omega_{ik}(t'-t'')}
\left[g^{\ast}_{\mu ,pk}(\mathbf{r}_{A},\mathbf{s}',\omega')
\hat{\sigma}_{ip}(t'') -
g^{\ast}_{\mu ,ip}(\mathbf{r}_{A},\mathbf{s}',\omega ')
\hat{\sigma}_{pk}(t'')\right]
\hat{f}^{\dagger}_{\mu}(\mathbf{s}',\omega')
\right)
\\ 
-g^{\ast}_{\lambda,ik}(\mathbf{r}_{A},\mathbf{s},\omega) 
\left(\hat{\sigma}_{kj}(0)e^{i\omega_{kj}t'}
-ie^{i\omega_{kj}(t'-t'')}
\left[g^{\ast}_{\mu ,pj}(\mathbf{r}_{A},\mathbf{s}',\omega')
\hat{\sigma}_{kp}(t'')
- g^{\ast}_{\mu,kp}(\mathbf{r}_{A},\mathbf{s}',\omega') 
\hat{\sigma}_{pj}(t'')\right]
\hat{f}^{\dagger}_{\mu}(\mathbf{s}',\omega')
\right)\bigg\}
\hat{f}^{\dagger}_{\lambda}(\mathbf{s},\omega)
\\+\mbox{terms containing}\,\hat{f}_{\lambda}(\mathbf{s},\omega)\,
\mbox{and/or}\,\hat{f}_{\mu}(\mathbf{s}',\omega')
\bigg] \,.
\label{sigma2}
\end{multline}
\end{widetext}

It is worth momentarily digressing from the derivation to consider the 
consistency of Eq. \eqref{sigma2} with the coupled Eqs. \eqref{eomrad} 
and \eqref{eomatom}. Here we have formally solved the atomic equations
of motion and substituted the result into the equation of motion for
the electromagnetic field, thereby describing the effect of the atom
on the field. In order to complete the analysis, one also has to
study the backreaction of the field on the atom. As a result of this backreaction, the
bare atomic transition frequencies $\tilde{\omega}_{ij}$ become
complex-valued quantities, gaining a line width 
$\Gamma_{ij}$ and a level shift $\delta\omega_{ij}$
\begin{equation}
\omega_{ij} = \tilde{\omega}_{ij} + \delta\omega_{ij} + i\Gamma_{ij}.
\label{dressed}
\end{equation}
The derivation of this result and the explicit expressions for 
the the line shifts $\delta\omega_{ij}$ and the line widths
$\Gamma_{ij}$, in terms of the dyadic Green
function, is well documented in the
literature (see, e.g. Refs.~\cite{acta,vogelwelsch} for reviews). In 
reality this feature will not be critical in the following derivation, but it is
important when considering the causality properties of the result (see
Appendix \ref{sec:kkr}). However, for the consistency of Eq. \eqref{sigma2} with the original equations of motion, the
modified transition frequencies $\omega_{ij}$ in Eq. \eqref{dressed}
must be used in place of their bare counterparts. 

We now return to Eq.~(\ref{sigma2}). In order to solve the integrals a
number of approximations need to be made. Firstly we write the bosonic field
operators as the product of a rapidly oscillating function and a
slowly varying envelope function,
\begin{equation}
\label{rwa}
\hat{f}_{\lambda}(\mathbf{s},\omega, t) 
=\tilde{\hat{f}}_{\lambda}(\mathbf{s},\omega, t)
e^{-i\omega t}.
\end{equation}
Secondly we assume that the radiation is off-resonant with any of the
atomic transitions and hence the frequency of the radiation field and
those associated with the atomic transitions are significantly
different.

We now apply the rotating-wave approximation. The lhs of
Eq.~(\ref{sigma2}) evolves at a frequency $\omega''$. Terms 
on the rhs of Eq.~(\ref{sigma2}) contain contributions from 
the two field modes evolving at $\omega$ and $\omega'$ and
contributions from the atomic operators. We are interested in the
process of parametric down conversion where the frequencies of the
incoming and outgoing photons combine such that $\omega''=\omega'+\omega$.
Terms on the rhs whose frequencies of evolution deviate significantly
from the above condition will oscillate rapidly in comparison to 
the resonant terms. Hence, over long time periods these non-resonant terms
average to zero. Thus, we keep all terms on the rhs 
of Eq.~(\ref{sigma2}) that satisfy $\omega''=\omega'+\omega$.
As a result we neglect terms which contain the annihilation operator 
$\hat{f}_{\lambda}(\mathbf{s},\omega)$.

Furthermore, in order to obey $\omega''=\omega'+\omega$, none of the (far off-resonant)
atomic transition frequencies $\omega_{ij}$ can appear on the
rhs. Hence, we drop all off-diagonal atomic operators and
retain only those terms that contain diagonal atomic projection
operators $\hat{\sigma}_{ii}$. Lastly terms with
$\hat{\sigma}_{ij}(0)$ represent the free (undriven) motion of the
atom in the background field and hence are not of interest here. 

The physical motivation for these approximations comes 
from the nature of the parametric down conversion process itself. 
The condition $\omega''=\omega'+\omega$ is a statement of energy conservation.
Hence there is no energy available to drive atomic transitions and
thus the atom must stay in its initial state. Therefore, the off
diagonal atomic operators, which describe atomic transitions, cannot
contribute. 

After applying these simplifications and permutating some indices we find
\begin{widetext}
\begin{align}
\dot{\hat{f}}^{\dagger}_{\nu}(\mathbf{r},\omega'') &=
i\omega ''\hat{f}^{\dagger}_{\nu}(\mathbf{r},\omega '') +
i\int\limits_0^{t}dt'\int\limits_0^{t'}dt''\int d^{3}s\int d^{3}s'\int
d\omega\int d\omega'\sum_{ijk}\hat{\sigma}_{ii}
\nonumber\\  
\times &\left\{e^{i\omega_{ij}t}e^{i(\omega - \omega_{kj})t'}
e^{i(\omega'-\omega_{ik}) t''}
g^{\ast}_{\lambda ,kj}(\mathbf{r}_{A},\mathbf{s},\omega)
g^{\ast}_{\mu ,ik}(\mathbf{r}_{A},\mathbf{s}',\omega ')
g_{\nu ,ij}(\mathbf{r}_{A},\mathbf{r},\omega '')\right.
\nonumber\\
&\left. - e^{i\omega_{kj}t}e^{i(\omega - \omega_{ij})t'}
e^{i(\omega ' - \omega_{ki}) t''}
g^{\ast}_{\lambda ,ij}(\mathbf{r}_{A},\mathbf{s},\omega)
g^{\ast}_{\mu ,ki}(\mathbf{r}_{A},\mathbf{s}',\omega ')
g_{\nu ,kj}(\mathbf{r}_{A},\mathbf{r},\omega '')\right.
\nonumber\\
&\left. - e^{i\omega_{kj}t}e^{i(\omega-\omega_{ki})t'}
e^{i(\omega ' - \omega_{ij})t''}
g^{\ast}_{\lambda ,ki}(\mathbf{r}_{A},\mathbf{s},\omega)
g^{\ast}_{\mu ,ij}(\mathbf{r}_{A},\mathbf{s}',\omega ')
g_{\nu ,kj}(\mathbf{r}_{A},\mathbf{r},\omega '')\right.
\nonumber\\
&\left. + e^{i\omega_{ji}t}
e^{i(\omega - \omega_{jk})t'}e^{i(\omega ' - \omega_{ki})t''}
g^{\ast}_{\lambda ,jk}(\mathbf{r}_{A},\mathbf{s},\omega)
g^{\ast}_{\mu ,ki}(\mathbf{r}_{A},\mathbf{s}',\omega ')
g_{\nu ,ji}(\mathbf{r}_{A},\mathbf{r},\omega '')\right\}
\tilde{\hat{f}}^{\dagger}_{\lambda}(\mathbf{s},\omega)
\tilde{\hat{f}}^{\dagger}_{\mu}(\mathbf{s}',\omega ').
\label{feom}
\end{align}
\end{widetext}
We would now like to perform the time
integrals. Note here that the coupling constants
$g_{\lambda,ij}(\mathbf{r},\mathbf{s},\omega)$ are not functions of
time and hence can be taken out of the integral. Since the slowly
varying envelope of the field operator
$\tilde{\hat{f}}^{\dagger}_{\lambda}(\mathbf{s},\omega)$ is
approximately constant over the time periods of interest, it can also
be taken out of the integral. Lastly the atomic operators
$\hat{\sigma}_{ii}$ are the projection operators on to the energy
eigenstates of the atomic Hamiltonian and hence stationary under
evolution by the atomic Hamiltonian.
Thus, the first term in Eq.~(\ref{feom}) integrates to
\begin{align}
&\hspace*{-2mm}\int\limits_0^{t}dt'\int\limits_0^{t'}dt'' e^{i\omega_{ij}t}
e^{i(\omega - \omega_{kj})t'}e^{i(\omega ' - \omega_{ik}) t''}
\nonumber\\
&\hspace*{-2mm}=
\frac{e^{i(\omega-\omega_{ki})t}-e^{i\omega_{ij}t}}
{(\omega'-\omega_{ik})(\omega-\omega_{kj})}
- \frac{e^{i(\omega + \omega ') t} -
e^{i\omega_{ij}t}}{(\omega'-\omega_{ik})(\omega+\omega'-\omega_{ij})} . 
\end{align}
Using Eq.~(\ref{rwa}) we can recombine the rapidly varying part of the
bosonic operators with the slowly varying envelope to recover the full
time dependent operator 
\begin{align}
&\left[\frac{e^{i(\omega - \omega_{kj})t} - e^{i\omega_{ij}t}}
{(\omega'-\omega_{ik})(\omega - \omega_{kj})} 
- \frac{e^{i(\omega + \omega ') t} -
e^{i\omega_{ij}t}}{(\omega'-\omega_{ik})(\omega+\omega'-\omega_{ij})}
\right]
\nonumber\\
&\hspace*{41mm}\times \tilde{\hat{f}}^{\dagger}_{\lambda}(\mathbf{s},\omega)
\tilde{\hat{f}}^{\dagger}_{\mu}(\mathbf{s}',\omega')
\nonumber\\
&= -\frac{1}{(\omega'-\omega_{ik})(\omega+\omega'-\omega_{ij})}
\hat{f}^{\dagger}_{\lambda}(\mathbf{s},\omega)
\hat{f}^{\dagger}_{\mu}(\mathbf{s}',\omega '),
\end{align}
where we have neglected the rapidly oscillating terms since these will
again average to zero over long time periods. Integrating the other
terms in Eq.~(\ref{feom}) in a similar way gives
\begin{multline}
\dot{\hat{f}}^{\dagger}_{\nu}(\mathbf{r},\omega'')
= i\omega ''\hat{f}^{\dagger}_{\nu}(\mathbf{r},\omega'')
- i\int d^{3}s\int d^{3}s'\int d\omega\int d\omega'
\\ 
\times
\hat{K}_{\lambda\mu\nu}(\mathbf{r}_{A};\mathbf{s},\mathbf{s}',\mathbf{r};
\omega,\omega',\omega'')
\hat{f}^{\dagger}_{\lambda}(\mathbf{s},\omega)
\hat{f}^{\dagger}_{\mu}(\mathbf{s}',\omega'),
\end{multline}
where we defined the nonlinear coupling tensor operator
$\hat{K}_{\lambda\mu\nu}(\mathbf{r}_{A};\mathbf{s},\mathbf{s}',\mathbf{r};
\omega,\omega',\omega'')$
as
\begin{align}
&\hat{K}_{\lambda\mu\nu}(\mathbf{r}_{A};\mathbf{s},\mathbf{s}',\mathbf{r};
\omega,\omega',\omega'')
= \sum_{ijk}\hat{\sigma}_{ii} \nonumber \\
&
\times\left\{
\frac{
g^{\ast}_{\lambda ,kj}(\mathbf{r}_{A},\mathbf{s},\omega)
g^{\ast}_{\mu,ik}(\mathbf{r}_{A},\mathbf{s}',\omega ')
g_{\nu,ij}(\mathbf{r}_{A},\mathbf{r},\omega'')}
{(\omega'-\omega_{ik})(\omega+\omega'-\omega_{ij})}
\right.
\nonumber \\ &
- \frac{
g^{\ast}_{\lambda ,ij}(\mathbf{r}_{A},\mathbf{s},\omega)
g^{\ast}_{\mu ,ki}(\mathbf{r}_{A},\mathbf{s}',\omega ')
g_{\nu ,kj}(\mathbf{r}_{A},\mathbf{r},\omega '')}
{(\omega'-\omega_{ki})(\omega+\omega'-\omega_{kj})}
\nonumber\\
&- \frac{
g^{\ast}_{\lambda ,ki}(\mathbf{r}_{A},\mathbf{s},\omega)
g^{\ast}_{\mu ,ij}(\mathbf{r}_{A},\mathbf{s}',\omega ')
g_{\nu ,kj}(\mathbf{r}_{A},\mathbf{r},\omega '')}
{(\omega ' - \omega_{ij})(\omega + \omega ' - \omega_{kj})}
\nonumber \\
&
\left.
+ \frac{
g^{\ast}_{\lambda ,jk}(\mathbf{r}_{A},\mathbf{s},\omega)
g^{\ast}_{\mu ,ki}(\mathbf{r}_{A},\mathbf{s}',\omega ')
g_{\nu ,ji}(\mathbf{r}_{A},\mathbf{r},\omega '')}
{(\omega ' - \omega_{ki})(\omega + \omega ' - \omega_{ji})}\right\}.
\label{K}
\end{align}

It is now straightforward to write down an effective interaction
Hamiltonian that, via Heisenberg's equations of motion, generates the
correct dynamical equation for the bosonic field operators:
\begin{align}
&\hat{H}^\mathrm{eff}_\mathrm{int} =
- \hbar\int d^{3}r\int d^{3}s\int d^{3}s'
\int d\omega\int d\omega' \int d\omega''
\nonumber \\
&\times \hat{K}_{\lambda\mu\nu}
(\mathbf{r}_{A};\mathbf{s},\mathbf{s}',\mathbf{r};\omega,\omega',\omega'')
\hat{f}^{\dagger}_{\lambda}(\mathbf{s},\omega)
\hat{f}^{\dagger}_{\mu}(\mathbf{s}',\omega')
\hat{f}_{\nu}(\mathbf{r},\omega'')
\nonumber\\
&\hspace*{60mm}+ \mbox{h.c.}\,.
\label{heff}
\end{align}
It is evident that the dynamical evolution of the atomic quantities is
frozen out in this approximation. The nonlinear coupling tensor
operator depends solely on the projection operators
$\hat{\sigma}_{ii}$ on to the atomic eigenstates. Since their
evolution is now static, $\dot{\hat{\sigma}}_{ii}=0$, we can replace
them by their expectation values $\rho^{(0)}_{ii}$. Hence, the
effective interaction Hamiltonian (\ref{heff}) becomes a functional of
the dynamical variables of the quantized electromagnetic field alone.

Although the interaction part of the Hamiltonian in
(\ref{heff}) correctly describes this process at a
microscopic level in terms of bosonic operators, it is the macroscopic
description, in terms of electric fields, that is of practical
interest. The response of an atom to an applied electric
field is described in terms of a single atom susceptibility or 
polarizability. Although an essential part of the macroscopic description,
the polarizability is, in fact, a function of the microscopic
properties of the atom. Hence we can use this to relate the
microscopic effective Hamiltonian to an equivalent macroscopic
effective Hamiltonian. It can be shown (see Appendices \ref{sec:chi}
and \ref{sec:kkr}) that the causal second order nonlinear
polarizability can be written as 
\begin{align}
&\chi_{\alpha\beta\gamma}^{(2)}(\omega ,\omega') =
\frac{1}{(i\hbar)^2\varepsilon_0}
\sum_{ijk}\rho^{(0)}_{ii}
\nonumber \\
&\times\left[\frac{d_{\alpha ,jk}d_{\beta ,ki}d_{\gamma ,ij}}
{(\omega ' - \omega_{ik})(\omega + \omega ' - \omega_{ij})}
- \frac{d_{\alpha ,ik}d_{\beta ,ji}d_{\gamma ,kj}}
{(\omega ' - \omega_{ij})(\omega + \omega ' - \omega_{kj})}\right.
\nonumber \\
&\left.
-\frac{d_{\alpha ,ji}d_{\beta ,ik}d_{\gamma ,kj}}
{(\omega ' - \omega_{ki})(\omega + \omega ' - \omega_{kj})}
+ \frac{d_{\alpha ,kj}d_{\beta ,ik}d_{\gamma ,ji}}
{(\omega ' - \omega_{ki})(\omega + \omega ' - \omega_{ji})}\right] .
\end{align}
One should also note that in using this
form of the polarizability requires the condition 
$\omega''=\omega+\omega'$. Using this we can write Eq.~(\ref{K}) in
terms of the second order nonlinear polarizability
\begin{align}
K_{\lambda\mu\nu}&(\mathbf{r}_{A};\mathbf{s},\mathbf{s}',\mathbf{r};
\omega,\omega',\omega'')
= \frac{i}{\hbar}\chi_{\alpha\beta\gamma}^{(2)}(\omega ,\omega ')
\left(\frac{\hbar\varepsilon_0}{\pi}\right)^{\frac{3}{2}}
\nonumber\\
&\times\frac{\omega^{2}\omega '^{2}\omega''^{2}}{c^6\varepsilon_0^2} 
\sqrt{\varepsilon''(\mathbf{s},\omega)\varepsilon''(\mathbf{s}',\omega')
\varepsilon''(\mathbf{r},\omega'')}
\nonumber \\
&\times G^{\ast}_{\alpha\lambda}(\mathbf{r}_{A},\mathbf{s},\omega)
G^{\ast}_{\beta\mu}(\mathbf{r}_{A},\mathbf{s}',\omega')
G_{\gamma\nu}(\mathbf{r}_{A},\mathbf{r},\omega'').
\end{align}
Hence the interaction term of the effective Hamiltonian becomes
\begin{align}
\hat{H}_\mathrm{int}^\mathrm{eff} &= -i\int\!d^3r\int\!d^3s\int\!d^3s'
\int\!d\omega\int\!d\omega' \int\!d\omega ''
\left(\frac{\hbar\varepsilon_0}{\pi}\right)^{\frac{3}{2}}
\nonumber \\
&\times
\chi_{\alpha\beta\gamma}^{(2)}(\omega ,\omega ')
\frac{\omega^{2}\omega'^{2}\omega''^{2}}{c^6\varepsilon_0^2}
\sqrt{\varepsilon''(\mathbf{s},\omega)
\varepsilon''(\mathbf{s}',\omega ')\varepsilon''(\mathbf{r},\omega'')} 
\nonumber\\
&\times G^{\ast}_{\alpha\lambda}(\mathbf{r}_{A},\mathbf{s},\omega)
G^{\ast}_{\beta\mu}(\mathbf{r}_{A},\mathbf{s}',\omega')
G_{\gamma\nu}(\mathbf{r}_{A},\mathbf{r},\omega '')
\nonumber\\
&\times\hat{f}^{\dagger}_{\lambda}(\mathbf{s},\omega)
\hat{f}^{\dagger}_{\mu}(\mathbf{s}',\omega ')
\hat{f}_{\nu}(\mathbf{r},\omega'') + \mbox{h.c.}\,.
\end{align}
Finally, we can combine the Green functions with the various factors
to re-form electric field operators
\begin{multline}
\hat{H}_\mathrm{int}^\mathrm{eff} = \varepsilon_0
\int d\omega\int d\omega' \int d\omega''
\chi_{\alpha\beta\gamma}^{(2)}(\omega,\omega')\\
\times\hat{E}^{\dagger}_{\alpha}(\mathbf{r}_{A},\omega)
\hat{E}^{\dagger}_{\beta}(\mathbf{r}_{A},\omega ')
\hat{E}_{\gamma}(\mathbf{r}_{A},\omega '') + \mbox{h.c.}\,.
\label{heffE}
\end{multline}
The Hamiltonian (\ref{heffE}) describes a second order nonlinear
interaction between the quantized electromagnetic field and a single
atom in free space, possibly near (but outside) a dielectric body.
In order to describe the situation in which the nonlinearly responding
atom is located inside a dielectric or even part of the dielectric
medium itself, an additional ingredient is necessary.

\section{Local Field Corrections}
\label{sec:lfc}

Previously, we have considered the nonlinear interaction of an
electric field and a single atom in free space. In this case the
applied fields act directly on the atom, and thus the local field at
the atom $\hat{E}_{\alpha}^\mathrm{loc}(\mathbf{r}_{A})$ is equal to the
applied field $\hat{E}_{\alpha}(\mathbf{r}_{A})$. In the case where
the interacting atom is part of a larger body, the electric field
at the position the atom is different from the applied external
field. The surrounding material modifies the applied field such that
\begin{equation}
\hat{E}_{\alpha}^\mathrm{loc}(\mathbf{r}_{A})
= \mathcal{L}[\varepsilon(\omega)]\hat{E}_{\alpha}(\mathbf{r}_{A}).
\end{equation}
The local field correction method involves calculating the prefactor
$\mathcal{L}[\varepsilon(\omega)]$ so that the local interaction can
be related to the applied fields. This is a common technique in linear 
optics and has even been applied to nonlinear processes \cite{sipe}. 
There are a number of ways to
perform these corrections. Here we shall consider the real
cavity model, which was first discussed in the framework of quantum
optics in Ref.~\cite{rcm} and in its present form in
Ref.~\cite{rcm2}. This technique is well known and has already been
used to calculate a number of atomic properties such as modified
spontaneous decay rates \cite{lfcdecay} and one and two atom van der
Waals interactions \cite{lfcvderw}.

In this model the interacting atom
is placed inside an empty spherical cavity of radius $R_{c}$, which
itself is embedded in the host medium. The local field correction is
performed by replacing the Green function found in the expansion of
the electric field with that of the spherical cavity. The Green
function for the spherical cavity can be found by considering wave
propagation from the cavity centred at $\mathbf{r}_{A}$ to a point
$\mathbf{r}$ located in the host medium. This is similar to the 
Onsager model \cite{onsager,onsager2} for local field corrections
(Appendix~\ref{ons}), a technique that is more common in classical
nonlinear optics, where the corrections to the classical fields are
calculated using similar concepts. The Green function method, however,
is more general as it takes into account the absorptive properties of
the surrounding material whereas the Onsager model does not. 
\begin{figure}
\includegraphics[width=0.6\linewidth]{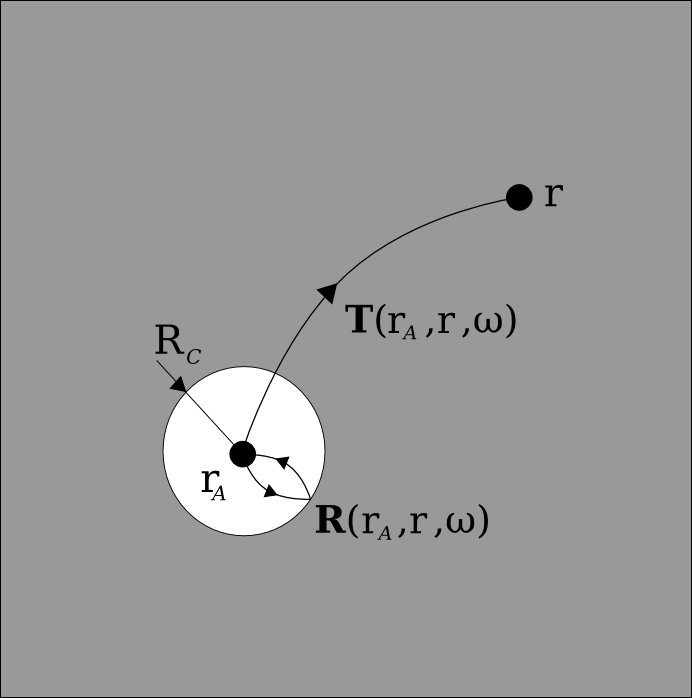}
\caption{The interacting atom is placed at the centre of a cavity of
radius $R_{c}$, embedded in an infinite homogeneous medium. The
scattering part of the Green function can be split into 
a contribution $R_{\alpha\beta}(\mathbf{r}_{A},\mathbf{r},\omega)$
related to reflection off the cavity wall and a contribution
$T_{\alpha\beta}(\mathbf{r}_{A},\mathbf{r},\omega)$ related to
transmission through the cavity wall. \label{Cavity}} 
\end{figure}

Consider an atom at the centre of an empty spherical cavity of radius
$R_{c}$ embedded in an infinite medium of permittivity
$\varepsilon(\omega)$ such that 
\begin{align}
\varepsilon(\mathbf{r},\omega) = \left\{
\begin{array}{lll}
1  & \mbox{if} & |\mathbf{r} - \mathbf{r}_{A}| < R_{c}\\
\varepsilon(\omega)  & \mbox{if} &
|\mathbf{r} - \mathbf{r}_{A}| \geq R_{c}
\end{array}
\right.
\end{align}
with $R_{c}$ on the order of the interatomic distance. The Green
function for the spherical cavity can be split up into two parts 
\begin{equation}
G_{\alpha\beta}(\mathbf{r}_{A},\mathbf{r},\omega) =
G_{\alpha\beta}^{C}(\mathbf{r}_{A},\mathbf{r},\omega) +
G_{\alpha\beta}^{S}(\mathbf{r}_{A},\mathbf{r},\omega), 
\end{equation}
where $G_{\alpha\beta}^{C}(\mathbf{r}_{A},\mathbf{r},\omega)$ is the
part which describes transmission within the cavity medium (i.e. free 
space) and $G_{\alpha\beta}^{S}(\mathbf{r}_{A},\mathbf{r},\omega)$ is
the part which describes scattering off the cavity wall. Furthermore,
we use the decomposition
\begin{equation}
G_{\alpha\beta}^{S}(\mathbf{r}_{A},\mathbf{r},\omega) =
R_{\alpha\beta}(\mathbf{r}_{A},\mathbf{r},\omega) +
T_{\alpha\beta}(\mathbf{r}_{A},\mathbf{r},\omega)
\end{equation}
where $R_{\alpha\beta}(\mathbf{r}_{A},\mathbf{r},\omega)$ is the
contribution from reflection off the cavity wall and
$T_{\alpha\beta}(\mathbf{r}_{A},\mathbf{r},\omega)$ is the contribution
from transmission through the cavity wall. We shall consider a
coarse-grained model with the characteristic length scale much greater
than $R_{c}$, the interatomic distance. Thus the individual atoms
cannot be resolved and so the medium can be considered to be a uniform
and continuous. Furthermore, since the cavity is of radius
$R_{c}$ there is only one resolvable point within the cavity, the
location of the atom, $\mathbf{r}_{A}$. As a result there can be no
propagation within the cavity and thus
$G_{\alpha\beta}^{C}(\mathbf{r}_{A},\mathbf{r},\omega)$ can be neglected.

The reflective part of the Green function is
\cite{lfcdecay,lfcvderw,lfcgreen}
\begin{equation}
R_{\alpha\beta}(\mathbf{r}_{A},\mathbf{r},\omega) =
\frac{i\omega}{6\pi}C(\omega)\frac{4}{3}\pi
R_{c}^{3}\delta(\mathbf{r}_{A} - \mathbf{r})\delta_{\alpha\beta}
\end{equation} 
with the Mie reflection coefficient
\begin{equation}
C(\omega) = \frac{h_1^{(1)}(z_0)\left[zh_1^{(1)}(z)\right]'-
\varepsilon(\omega)h_1^{(1)}(z) \left[z_0h_1^{(1)}(z_0)\right]'}
{\varepsilon(\omega)h_1^{(1)}(z) \left[z_0j_1^{(1)}(z_0)\right]' -
j_1^{(1)}(z_0)\left[zh_1^{(1)}(z)\right]'} 
\end{equation}
where $z_0=\omega R_{c}/c$ and
$z=\sqrt{\varepsilon(\omega)}\omega R_{c}/c$ and
$j_1^{(1)}(z)$ and $h_1^{(1)}(z)$ are, respectively, the first
spherical Bessel and Hankel functions of the first kind,
\begin{equation}
j_1^{(1)}(z) = \frac{\sin(z)}{z^2} - \frac{\cos(z)}{z},\quad
h_1^{(1)}(z) = \left(\frac{1}{z} + \frac{i}{z^2}\right)e^{iz}.
\end{equation}
As $R_{c}$ is small compared to the optical wavelengths associated
with the process (i.e. $R_{c}\ll c/\omega$) we can expand $C(\omega)$
in powers of $\omega R_{c}/c$
\begin{align}
C(\omega) &= 3\frac{\varepsilon(\omega)-1}{[2\varepsilon(\omega)+1]}
\frac{c^{3}}{i\omega^{3}R_{c}^{3}}
\nonumber\\
&\quad+\frac{9}{5}\left\{
\frac{[\varepsilon(\omega)-1][4\varepsilon(\omega)+1]}
{[2\varepsilon(\omega)+1]^2}\right\}
\frac{c}{i\omega R_{c}}
\nonumber\\ 
&\qquad + 9\frac{\varepsilon(\omega)n^{3}(\omega)}
{[2\varepsilon(\omega)+1]^2} - 1
+ \mathcal{O}\left(\frac{\omega R_{c}}{c}\right).
\end{align}
Substituting these expressions into
$R_{\alpha\beta}(\mathbf{r}_{A},\mathbf{r},\omega)$ and taking the
cavity radius to zero ($\omega R_{c}/c\rightarrow 0$) gives 
\begin{equation}
R_{\alpha\beta}(\mathbf{r}_{A},\mathbf{r},\omega) =
\frac{2}{3}\frac{\varepsilon(\omega) - 1}{2\varepsilon(\omega)+1}
\frac{c^2}{\omega^2}\delta(\mathbf{r}_{A} - \mathbf{r})\delta_{\alpha\beta}. 
\end{equation}

The transmission part of the Green function is
\cite{lfcdecay,lfcvderw,lfcgreen}
\begin{equation}
T_{\alpha\beta}(\mathbf{r}_{A},\mathbf{r},\omega) =
D(\omega)G_{\alpha\beta}^{B}(\mathbf{r}_{A},\mathbf{r},\omega) 
\end{equation}
with the Mie transmission coefficient
\begin{equation}
D(\omega) = \frac{j_1^{(1)}(z_0)\left[z_0h_1^{(1)}(z_0)\right]'
- h_1^{(1)}(z_0)\left[z_0j_1^{(1)}(z_0)\right]'}
{j_1^{(1)}(z_0)\left[zh_1^{(1)}(z)\right]'
- \varepsilon(\omega)h_1^{(1)}(z)\left[z_0j_1^{(1)}(z_0)\right]'}.
\end{equation}
Here, $G_{\alpha\beta}^\mathrm{B}(\mathbf{r}_{A},\mathbf{r},\omega)$
is the bulk Green function for the infinitely extended medium without
cavities. Expanding $D(\omega)$ in powers of $\omega R_{c}/c$ gives
\begin{equation}
D(\omega) = \frac{3\varepsilon(\omega)}{2\varepsilon(\omega)+1} +
\mathcal{O}\left(\frac{\omega R_{c}}{c}\right)
\end{equation}
which, in the limit $\omega R_{c}/c\rightarrow 0$, leads to
\begin{equation}
T_{\alpha\beta}(\mathbf{r}_{A},\mathbf{r},\omega) =
\frac{3\varepsilon(\omega)}{2\varepsilon(\omega)+1}
G_{\alpha\beta}^\mathrm{B}(\mathbf{r}_{A},\mathbf{r},\omega) 
\end{equation}
for the transmission part of the Green function. Collecting all the
results yields
\begin{align}
G_{\alpha\beta}(\mathbf{r}_{A},\mathbf{r},\omega) &=
\frac{2}{3}\frac{\varepsilon(\omega)-1}{2\varepsilon(\omega)+1}
\frac{c^2}{\omega^2}\delta(\mathbf{r}_{A}-\mathbf{r})\delta_{\alpha\beta}
\nonumber\\ 
&\quad + \frac{3\varepsilon(\omega)}{2\varepsilon(\omega)+1}
G_{\alpha\beta}^\mathrm{B}(\mathbf{r}_{A},\mathbf{r},\omega)
\label{lfcgreen}
\end{align}
for the local-field corrected Green function including the cavity. 
One should note that the $\delta$-function in Eq. \eqref{lfcgreen} 
is the contribution from the reflective part of the scattering Green function. 
The intrinsic singularity present in the bulk Green function is still contained
within $G_{\alpha\beta}^\mathrm{B}(\mathbf{r}_{A},\mathbf{r},\omega)$.

Here we have used the Green function for a cavity embedded in an infinite 
homogeneous media to locally field correct the electric field at the location of the 
atom. However, it was shown in Ref. \cite{lfcdecay} that this result can be easily 
generalized to cavities embedded in arbitrary media. To obtain the expression 
for arbitrary media one merely needs to replace the bulk Green function and 
the homogeneous permittivity with the appropriate Green function and 
inhomogeneous permittivity for the new geometry. Hence the following results 
are, with the right substitutions, correct for any system.

We can now apply these local-field corrections to the nonlinear
Hamiltonian. Inserting the local-field corrected Green function 
(\ref{lfcgreen}) into the effective nonlinear Hamiltonian
(\ref{heffE}) gives
\begin{align}
&\hat{H}_\mathrm{int}^\mathrm{eff} = i
\int\!d^3r\int\!d^3r'\int\!d^3r''\int\!d\omega \int\!d\omega'\int\!d\omega''
\left(\frac{\hbar\varepsilon_0}{\pi}\right)^{\frac{3}{2}}
\nonumber \\ &\times
\chi_{\alpha\beta\gamma}^{(2)}(\omega ,\omega ')
\frac{\omega^2\omega'^{2}\omega''^{2}}{c^6\varepsilon_0^2}
\sqrt{\varepsilon''(\mathbf{s},\omega)
\varepsilon''(\mathbf{s}',\omega')\varepsilon''(\mathbf{r},\omega'')}
\nonumber\\
&\times\left[
\tilde{C}^{\ast}(\omega )\frac{c^2}{\omega^2}
\delta(\mathbf{r}_{A} - \mathbf{r})\delta_{\alpha\lambda}
+ \tilde{D}^{\ast}(\omega)
G_{\alpha\lambda}^\mathrm{B\ast}(\mathbf{r}_{A},\mathbf{r},\omega)\right]
\nonumber\\
&\times\left[
\tilde{C}^{\ast}(\omega ')\frac{c^2}{\omega'^{2}}
\delta(\mathbf{r}_{A}-\mathbf{r}')\delta_{\beta\mu}
+ \tilde{D}^{\ast}(\omega ')
G_{\beta\mu}^\mathrm{B\ast}(\mathbf{r}_{A},\mathbf{r}',\omega ')\right]
\nonumber\\
&\times\left[
\tilde{C}(\omega '')\frac{c^2}{\omega ''^{2}}
\delta(\mathbf{r}_{A} - \mathbf{r}'')\delta_{\gamma\nu}
+ \tilde{D}(\omega '')
G_{\gamma\nu}^\mathrm{B}(\mathbf{r}_{A},\mathbf{r}'',\omega '')\right]
\nonumber \\
&\hspace*{10mm}\times\hat{f}_{\lambda}^{\dagger}(\mathbf{r},\omega)
\hat{f}^{\dagger}_{\mu}(\mathbf{r}',\omega ')
\hat{f}_{\nu}(\mathbf{r}'',\omega '')  + \mbox{h.c.}
\end{align}
where we abbreviated
\begin{equation}
\tilde{C}(\omega) = \frac{2}{3}
\frac{\varepsilon(\omega)-1}{2\varepsilon(\omega)+1}
\,,\quad
\tilde{D}(\omega) =\frac{3\varepsilon(\omega)}{2\varepsilon(\omega)+1} \,.
\end{equation}

Recall that $\chi_{\alpha\beta\gamma}^{(2)}(\omega,\omega')$ is the 
polarizability for an isolated atom in free space and that
$\omega''=\omega+\omega' $. Inside a material body, the interpretation
of the nonlinear polarizability has changed. We can thus define a
local-field corrected polarizability  (i.e. the polarizability of an atom embedded in an
extended medium) as  
\begin{equation}
\tilde{\chi}_{\alpha\beta\gamma}^{(2)}(\omega ,\omega ') =
\tilde{D}^{\ast}(\omega)\tilde{D}^{\ast}(\omega ')\tilde{D}(\omega '')
\chi_{\alpha\beta\gamma}^{(2)}(\omega,\omega').
\end{equation}
Performing the integrations over the $\delta$-functions and recombining
the various factors to form electric and linear noise polarization
fields gives
\begin{align}
\hat{H}_\mathrm{int}^\mathrm{eff} &= \varepsilon_0
\int d\omega \int d\omega'\int d\omega''
\tilde{\chi}_{\alpha\beta\gamma}^{(2)}(\omega ,\omega ')
\nonumber\\ 
&\times
\left[ \hat{E}_{\alpha}^{\dagger}(\mathbf{r}_{A},\omega) +
\mathcal{L}[\varepsilon^\ast(\omega)] 
\hat{P}^{\dagger}_{\mathrm{N},\alpha}(\mathbf{r}_{A},\omega)\right]
\nonumber\\ 
&\times
\left[ \hat{E}^{\dagger}_{\beta}(\mathbf{r}_{A},\omega ')
+\mathcal{L}[\varepsilon^{\ast}(\omega ')]
\hat{P}^{\dagger}_{\mathrm{N},\beta}(\mathbf{r}_{A},\omega ')
\right]
\nonumber\\ 
&\times
\left[ \hat{E}_{\gamma}(\mathbf{r}_{A},\omega'')
+\mathcal{L}[\varepsilon(\omega'')]
\hat{P}_{\mathrm{N},\gamma}(\mathbf{r}_{A},\omega'')
\right] + \mbox{h.c.}
\label{finalH}
\end{align}
where
\begin{eqnarray}
\mathcal{L}[\varepsilon(\omega)] = \frac{2}{9\varepsilon_0}
\frac{\varepsilon(\omega) - 1}{\varepsilon(\omega)}
\end{eqnarray}
is the local correction factor for the noise polarization field.

The interaction Hamiltonian (\ref{finalH}) is trilinear in the electric 
field and the (local-field corrected) noise polarization field with the 
strength of the interaction characterised by the (local-field corrected) 
second order polarizability
$\tilde{\chi}_{\alpha\beta\gamma}^{(2)}(\omega,\omega')$. This can be 
viewed as the combination of several different types of interaction
processes. The term trilinear in the electric field corresponds to the
sought parametric down-conversion process. All other terms
describe various nonlinear interactions between the electric field 
and noise polarization fields. They correspond to the absorption of
one or both of the outgoing photons by the medium, the production of
one or two outgoing photons from an excited noise field (e.g. by
thermal excitation) and a pure nonlinear noise field
interaction. These extra interaction terms are features of the
absorptive properties of the medium and are in effect corrections to
lower-order (nonlinear) processes. 

On taking the limit of vanishing absorption the noise polarization fields 
vanish identically [recall Eq.~(\ref{eq:noisepol})]. Hence, the interaction
Hamiltonian (\ref{finalH}) reduces to the standard interaction Hamiltonian 
associated with parametric down conversion in non-absorbing media. It should 
be noted that computing higher-order nonlinear processes will similarly 
lead to corrections to second order processes.  However, in view of the 
applied rotating-wave approximation, their effect remains negligible.

\section{Summary}
\label{sec:sum}

Beginning with the total Hamiltonian for the medium-assisted 
electromagnetic field interaction with a single $N$-level atom 
(\ref{eq:linearH}), we derived Heisenberg's equations of motion 
for the dynamical variables of both the atom and the medium-assisted field. By 
integrating out the atomic degrees of freedom and then focusing 
on one particular term in the nonlinear expansion of the field 
variables, we have found an effective equation of motion for one 
particular optical process. One can think of this equation of motion 
as the dynamical equation that is generated from an effective 
Hamiltonian which describes only the process of interest. This 
effective Hamiltonian does not refer to the $N$-level atom, 
consisting only of the dynamical variables relating to the 
medium-assisted field. All information about the effect of the atom 
is now contained in the coupling constant for the interaction terms.

Here, as a specific example, by studying the equations of motion to 
second order in perturbation theory we have derived an effective 
Hamiltonian for the second order nonlinear process of parametric 
down conversion in the presence of an absorbing host material. By 
applying local-field corrections to the Hamiltonian we have moved 
from considering the interaction of light with a single atom to 
considering the interaction of light with a bulk material. We found 
that the Hamiltonian can be expressed in terms of various products 
of electric and linear noise polarization fields.

The local-field corrections have two effects: to replace the free-space
second order nonlinear polarizability
$\chi_{\alpha\beta\gamma}^{(2)}(\omega,\omega')$ by its bulk modified
counterpart $\tilde{\chi}_{\alpha\beta\gamma}^{(2)}(\omega,\omega')$,
and to introduce additional contributions to the effective interaction
Hamiltonian. These additional terms are related to nonlinear absorption
processes inside the bulk material, similar to those found in the
purely macroscopic approach pursued in
\cite{nonlinear1,nonlinear2}. In contrast to the macroscopic picture,
in our present microscopic derivation we have gained a better
understanding of the origins of the additional contributions to the
effective Hamiltonian. 

In the limit of vanishing absorption, where the noise polarization field 
disappears, one recovers the standard second order effective interaction
Hamiltonian as used in classical nonlinear optics. 
In the generic situation when absorption cannot be disregarded, the 
effective Hamiltonian (\ref{finalH}) will be the starting point for 
subsequent investigations into the role of absorption on the generation
of down-converted photons and their propagation through nonlinear media.

\section{Acknowledgements}
\label{sec:ack}

This work was supported by the UK Engineering and Physical Sciences
Research Council.

\appendix

\section{Second order polarizability}
\label{sec:chi}

Here we derive the second order susceptibility of a single atom 
(also known as the second order polarizability) in terms of the microscopic properties 
of the atom. A semi-classical approach with classical radiation and a 
quantized atom is used. This approach follows closely the derivation in
Ref.~\cite{schubertwilhelmi}.

In general media the susceptibility $\bm{\chi}$ is a tensorial function that relates the
polarization of the medium to the strength of the applied electric
field. In the time domain, for the case of second order processes, the
polarization is related to the applied electric field by
\begin{align}
&P_{\gamma}^{(2)}(\mathbf{r},t) = \nonumber \\
&\varepsilon_0\int\limits_0^{\infty}d\tau
\int\limits_{\tau}^{\infty}d\tau '\chi_{\alpha\beta\gamma}^{(2)}(\tau,\tau ')
E_{\alpha}(\mathbf{r},t - \tau)E_{\beta}(\mathbf{r},t - \tau ').
\label{polarization}
\end{align}
Here $\chi_{\alpha\beta\gamma}^{(2)}$ is the second order susceptibility. 
If the medium consists of a single atom, Eq. \eqref{polarization} still applies 
with the left hand side giving the polarization of the single atom. The susceptibility 
that appears in the single atom version of Eq. \eqref{polarization} is often 
referred to as the second order polarizability of the atom.

Consider a single atom in state $\hat{\rho}$ which evolves under the
perturbed Hamiltonian
$\hat{H}=\hat{H}_\mathrm{A}+\hat{H}_\mathrm{int}$. Since the operator
$\hat{\rho}$ can be written as a sum over the projection operators
onto the energy eigenstates of the atomic Hamiltonian, the state is
time stationary under evolution of the unperturbed atomic
Hamiltonian $\hat{H}_\mathrm{A}$. Hence the Heisenberg equation of
motion for the atom is
\begin{equation}
\frac{\partial\hat{\rho}(t)}{\partial t} =
\frac{1}{i\hbar}[\hat{\rho}(t),\hat{H}_\mathrm{int}(t)].
\end{equation}
This can be formally solved to give the recursive relation
\begin{equation}
\hat{\rho}(t) = \frac{1}{i\hbar}\int\limits^t_{t_0}dt_1
[\hat{\rho}(t_1),\hat{H}_\mathrm{int}(t_1)] + \rho^{(0)}
\label{eq:rho}
\end{equation}
where $\rho^{(0)}$ is a constant of integration and is equal to the
state of the system at $t=t_0$. Equation~(\ref{eq:rho}) can be
resubstituted into itself to give a series solution for $\hat{\rho}(t)$, 
\begin{multline}
\hat{\rho}(t) = \sum_{n=1}^{\infty}\frac{1}{(i\hbar)^{n}}
\int\limits_{t_0}^{t}dt_1 \ldots \int\limits_{t_0}^{t_{n-1}}dt_{n} \nonumber\\
\times\left[[\cdots[\hat{\rho}^{(0)},\hat{H}_\mathrm{int}(t_{n})]
\cdots ,\hat{H}_\mathrm{int}(t_2)],\hat{H}_\mathrm{int}(t_1)\right] +
\rho^{(0)}. 
\end{multline}
Given that we now have an expression for the quantum state of the
atom, we can now write down the polarization of the  atom, which is defined as
the expectation value of the dipole moment operator 
\begin{equation}
P_{\gamma}(t) = \langle\hat{d}_{\gamma}\rangle =
\mathrm{Tr}\left[\hat{\rho}(t)\hat{d}_{\gamma}(t)\right]. 
\end{equation}
As $\hat{\rho}(t)$ is an infinite expansion so $P_{\gamma}(t)$ will
be an infinite expansion. Here second order processes are of interest
so only the second order term is considered, 
\begin{align}
&P_{\gamma}^{(2)}(t) = \nonumber \\
&\frac{1}{(i\hbar)^2}\int\limits_{t_0}^t dt_1\int\limits_{t_0}^{t_1}dt_2
\mathrm{Tr}\left\{\left[[\hat{\rho}^{(0)},\hat{H}_\mathrm{int}(t_2)],
\hat{H}_\mathrm{int}(t_1)\right]\hat{d}_{\gamma}(t)\right\}.
\label{eq:p}
\end{align}
In the dipole approximation, $\hat{H}_\mathrm{int}(t)$ is given by
\begin{equation}
\hat{H}_\mathrm{int}(t) = - \hat{d}_{\alpha}(t)E_{\alpha}(t).
\end{equation}
Hence (\ref{eq:p}) becomes
\begin{align}
&P_{\gamma}^{(2)}(t) = \frac{1}{(i\hbar)^2}
\int\limits_{t_0}^{t}dt_1\int\limits_{t_0}^{t_1}dt_2 \nonumber\\
&\quad\times \mathrm{Tr}\left\{\left[[\hat{\rho}^{(0)},
-\hat{d}_{\beta}(t_2)E_{\beta}(t_2)],
-\hat{d}_{\alpha}(t_1)E_{\alpha}(t_1)\right]
\hat{d}_{\gamma}(t)\right\}
\nonumber\\
&= \frac{1}{(i\hbar)^2}
\int\limits_{t_0}^{t}dt_1\int\limits_{t_0}^{t_1}dt_2\nonumber\\
&\quad\times \mathrm{Tr}\left\{\hat{\rho}^{(0)}\left[
[\hat{d}_{\gamma}(t),\hat{d}_{\alpha}(t_1)],\hat{d}_{\beta}(t_2)\right]\right\}
E_{\alpha}(t_1)E_{\beta}(t_2),
\label{eq:P1}
\end{align}
where the cyclicity of the trace and the classical nature of the
radiation have been used to arrive at the result. We now make the
substitutions $t_1=t-\tau_1$ and $t_2=t-\tau_2$
and take $t_0 \rightarrow -\infty$ (and hence
$\tau_{1,2} \rightarrow = +\infty$). Hence Eq.~(\ref{eq:P1}) becomes
\begin{multline}
P_{\gamma}^{(2)}(t) = \frac{1}{(i\hbar)^2}
\int\limits_{\infty}^{0}d\tau_1\int\limits_{\infty}^{\tau_1}d\tau_2
\\
\times \mathrm{Tr}\left\{\hat{\rho}^{(0)}\left[
[\hat{d}_{\gamma}(t),\hat{d}_{\alpha}(t - \tau_1)],\hat{d}_{\beta}(t - \tau_2)\right]
\right\}\\
\times E_{\alpha}(t - \tau_1)E_{\beta}(t - \tau_2).
\label{eq:P2}
\end{multline}
The time dependence of quantum operators can be written as
$\hat{O}(t) = \hat{U}^{\dagger}(t)\hat{O}\hat{U}(t)$
where $\hat{U}$ is the unitary operator
$\hat{U} = e^{-i\hat{H}t/\hbar}$ associated with the Hamiltonian $\hat{H}$.
Hence, we can write 
\begin{equation}
\hat{d}_{\alpha}(t - \tau) = \hat{U}^{\dagger}(t)
\hat{d}_{\alpha}(-\tau)\hat{U}(t). 
\end{equation}
Noting that the dipole moment operator commutes with
$\hat{H}_\mathrm{int}$ and that $\rho^{(0)}$ commutes with
$\hat{H}_\mathrm{A}$ and again using the cyclicity of the trace,
Eq.~(\ref{eq:P2}) becomes
\begin{multline}
P_{\gamma}^{(2)}(t) =
\frac{1}{(i\hbar)^2}
\int\limits^{\infty}_0d\tau_1\int\limits^{\infty}_{\tau_1}d\tau_2
 \\ 
\times\mathrm{Tr}\left\{\hat{\rho}^{(0)}\left[
[\hat{d}_{\gamma}(0),\hat{d}_{\alpha}(-\tau_1)],
\hat{d}_{\beta}(-\tau_2)\right]\right\}\\
\times E_{\alpha}(t - \tau_1)E_{\beta}(t - \tau_2).
\label{eq:P3}
\end{multline}
Comparing Eq. (\ref{eq:P3}) with Eq. (\ref{polarization}) one can see
that the second order nonlinear polarizability is given by 
\begin{multline}
\chi_{\alpha\beta\gamma}^{(2)}(\tau_1,\tau_2) = \\
\frac{1}{(i\hbar)^2}
\frac{1}{\varepsilon_0} \mathrm{Tr}\left\{\hat{\rho}^{(0)}
\left[[\hat{d}_{\gamma}(0),\hat{d}_{\alpha}( - \tau_1)],
\hat{d}_{\beta}( - \tau_2)\right]\right\}.\\
\end{multline}

In the frequency domain, the Fourier transforms of the electric field
and the polarization field are substituted into Eq. (\ref{eq:P3}), and
we find that
\begin{align}
&P_{\gamma}^{(2)}(\omega '') = \frac{1}{(i\hbar)^2}
\int\limits^{\infty}_0 d\tau_1 \int\limits^{\infty}_{\tau_1}d\tau_2
e^{-i\omega\tau_1}e^{-i\omega '\tau_2}  \nonumber\\
&\quad\times \mathrm{Tr}\left\{\hat{\rho}^{(0)}\left[
[\hat{d}_{\gamma}(0),\hat{d}_{\alpha}(-\tau_1)],
\hat{d}_{\beta}(-\tau_2)\right] \right\}
E_{\alpha}(\omega)E_{\beta}(\omega'),
\end{align}
and hence
\begin{multline}
\chi_{\alpha\beta\gamma}^{(2)}(\omega,\omega ') = \frac{1}{(i\hbar)^2}
\frac{1}{\varepsilon_0}\int\limits^{\infty}_0 d\tau_1
\int\limits^{\infty}_{\tau_1}d\tau_2 e^{-i\omega\tau_1}e^{-i\omega'\tau_2}
 \\
\times \mathrm{Tr}\left\{\hat{\rho}^{(0)}\left[
[\hat{d}_{\gamma}(0),\hat{d}_{\alpha}(-\tau_1)],
\hat{d}_{\beta}(-\tau_2)\right]\right\}.\\
\end{multline}
Note here that, by performing the Fourier transform, we have 
imposed the condition $\omega''=\omega+\omega'$.
The trace of the combination of dipole moment operators can be
computed by inserting identity operators in terms of atomic energy
eigenstates $|\epsilon_i\rangle$. The matrix elements of the dipole
moment operators are given by
\begin{multline}
\label{eq:trace1}
\langle\epsilon_{i}|\hat{d}_{\alpha}(t)|\epsilon_{j}\rangle =
\langle\epsilon_{i}|\hat{U}^{\dagger}(t)\hat{d}_{\alpha}\hat{U}(t)|
\epsilon_{j}\rangle
\\ 
= \langle\epsilon_{i}|e^{i\hat{H}_\mathrm{A}t/\hbar}\hat{d}_{\alpha}
e^{-i\hat{H}_\mathrm{A}t/\hbar}|\epsilon_{j}\rangle
= d_{\alpha, ij}e^{i\omega_{ij}t}.
\end{multline}
Here $\epsilon_{i}$ is the energy eigenvalue associated with the
eigenstate $|\epsilon_{i}\rangle$. Note also that $\hat{\rho}^{(0)}$
is diagonal in the eigenbasis of the atomic Hamiltonian
$\hat{H}_\mathrm{A}$,
\begin{equation}
\label{eq:trace2}
\langle\epsilon_{i}|\hat{\rho}^{(0)}|\epsilon_{j}\rangle =
\rho^{(0)}_{ij}\delta_{ij}.
\end{equation}
Using the relations (\ref{eq:trace1}), (\ref{eq:trace2}) and the
completeness relation for the atomic eigenstates, the second order
susceptibility becomes
\begin{align}
\chi_{\alpha\beta\gamma}^{(2)}&(\omega ,\omega ') =
\int\limits_0^{\infty}d\tau_1\int\limits_{\tau_1}^{\infty}d\tau_2
\frac{1}{(i\hbar)^2}\frac{1}{\varepsilon_0}
\sum_{ijk}\rho^{(0)}_{ii}
\nonumber \\ \times
&\left[d_{\alpha ,jk}d_{\beta ,ki}d_{\gamma ,ij}
e^{-i(\omega-\omega_{kj})\tau_1}e^{-i(\omega'-\omega_{ik})\tau_2}
\right.\nonumber \\
&\left. - d_{\alpha ,ij}d_{\beta ,ki}d_{\gamma ,jk}
e^{-i(\omega-\omega_{ji})\tau_1}e^{-i(\omega'-\omega_{ik})\tau_2}
\right.\nonumber \\
&\left. - d_{\alpha ,ki}d_{\beta ,ij}d_{\gamma ,jk}
e^{-i(\omega-\omega_{ik})\tau_1}e^{-i(\omega'-\omega_{ji})\tau_2}
\right.\nonumber \\
&\left. + d_{\alpha ,jk}d_{\beta ,ij}d_{\gamma ,ki}
e^{-i(\omega-\omega_{kj})\tau_1}e^{-i(\omega'-\omega_{ji})\tau_2}
\right].\nonumber \\
\end{align}
It is important to note that, in order for the expression
to be consistent with the atom-field equations of motion, the
transition frequencies $\omega_{ij}$ must be complex variables  
\begin{equation}
\omega_{ij} = \tilde{\omega}_{ij} + \delta\omega_{ij} + i\Gamma_{ij}.
\end{equation}
As in Eq.~(\ref{dressed}), $\tilde{\omega}_{ij}$ is the bare atomic
transition frequency, $\delta\omega_{ij}$ is the level shift and
$\Gamma_{ij}$ is the transition linewidth. This results in a factor of
$e^{-\Gamma\tau}$ in the integrand which leads to convergence at the
upper limit of the integral. Integrating both time integrals leads to
\begin{align}
&\chi_{\alpha\beta\gamma}^{(2)}(\omega ,\omega ') = \frac{1}{(i\hbar)^2}
\frac{1}{\varepsilon_0}\sum_{ijk}\rho^{(0)}_{ii}\nonumber \\
&\times\!\left[\frac{d_{\alpha ,jk}d_{\beta ,ki}d_{\gamma ,ij}}
{(\omega ' - \omega_{ik})(\omega + \omega ' - \omega_{ij})}
- \frac{d_{\alpha ,ij}d_{\beta ,ki}d_{\gamma ,jk}}
{(\omega'-\omega_{ik})(\omega+\omega'-\omega_{jk})}\right.\nonumber \\
&\left. -\frac{d_{\alpha ,ki}d_{\beta ,ij}d_{\gamma ,jk}}
{(\omega'-\omega_{ji})(\omega+\omega'-\omega_{jk})}
+ \frac{d_{\alpha ,jk}d_{\beta ,ij}d_{\gamma ,ki}}
{(\omega'-\omega_{ji})(\omega+\omega'-\omega_{ki})}\right]
\end{align}
which is the expression for the polarizability of a single atom in
frequency space. 

\section{Causality and the Kramers-Kronig Relations}
\label{sec:kkr}

The linear polarization field is a linear response to the applied
electric field. As with any response theory the magnitude of the
reaction is described by the response function. In the case of the linear
polarization field the response function is the linear susceptibility 
\begin{equation}
P_{\alpha}(\mathbf{r},t) = \varepsilon_0\int\limits_0^{\infty}d\tau\,
\chi_{\alpha\beta}^{(1)}(\tau)E_{\beta}(\mathbf{r},t-\tau).
\end{equation}
By causality $\chi_{\alpha\beta}^{(1)}(\tau)$ must vanish for $\tau<0$; the
polarization field at time $t$ cannot depend on electric fields at
times greater than $t$. Thus 
\begin{equation}
\chi_{\alpha\beta}^{(1)}(\tau) = \Theta(\tau)\chi_{\alpha\beta}^{(1)}(\tau),
\end{equation}
where $\Theta(\tau)$ is the Heaviside step function. Fourier
transforming both sides of the equation gives
\begin{equation}
\chi_{\alpha\beta}^{(1)}(\omega) =
\frac{1}{2\pi i}
\int\limits_{-\infty}^{\infty}d\omega'
\frac{\chi_{\alpha\beta}^{(1)}(\omega')}{\omega-\omega'}.
\end{equation}
The function $1/(\omega-\omega')$ has to be seen in its distributional
sense and, by using Sochotzki's formula, can be decomposed into
$1/(\omega-\omega')=\mathcal{P}/(\omega-\omega')+i\pi\delta/(\omega-\omega')$.
Here $\mathcal{P}$ denotes the principal part. This results in
\begin{equation}
\label{eq:kk1}
\chi_{\alpha\beta}^{(1)}(\omega) = \frac{\mathcal{P}}{\pi i}
\int\limits_{-\infty}^{\infty}d\omega'
\frac{\chi_{\alpha\beta}^{(1)}(\omega')}{\omega-\omega'}.
\end{equation}
Decomposing Eq.~(\ref{eq:kk1}) into its real and imaginary parts gives
\begin{align}
\mathrm{Re}\left[\chi_{\alpha\beta}^{(1)}(\omega)\right] &= \frac{\mathcal{P}}{\pi}
\int\limits_{-\infty}^{\infty}d\omega '
\frac{\mathrm{Im}\left[\chi_{\alpha\beta}^{(1)}(\omega')\right]}{(\omega-\omega')},
\\
\mathrm{Im}\left[\chi_{\alpha\beta}^{(1)}(\omega)\right] &= -\frac{\mathcal{P}}{\pi}
\int\limits_{-\infty}^{\infty}d\omega '\frac{\mathrm{Re}
\left[\chi_{\alpha\beta}^{(1)}(\omega')\right]}{(\omega-\omega')}.
\end{align}
These are the linear Kramers-Kronig relations. A causal linear
response function must satisfy these relations. For more information
about the linear Kramers-Kronig relations the reader is referred to
Ref. \cite{jackson}. 

One can derive a second order nonlinear version of the Kramers-Kronig
relations in a similar way. The second order nonlinear polarization
is given by 
\begin{align}
&P_{\alpha}^{(2)}(\mathbf{r},t) = \nonumber\\
&\varepsilon_0\int\limits_0^{\infty}d\tau
\int\limits_{\tau}^{\infty}d\tau'\, 
\chi_{\alpha\beta\gamma}^{(2)}(\tau,\tau')
E_{\beta}(\mathbf{r},t-\tau)E_{\gamma}(\mathbf{r},t-\tau'),
\end{align}
where $\chi_{\alpha\beta\gamma}^{(2)}(\tau,\tau')$ is the second order
response function. Owing to causality,
$\chi_{\alpha\beta\gamma}^{(2)}(\tau,\tau')$ must satisfy 
\begin{equation}
\label{eq:kk2}
\chi_{\alpha\beta\gamma}^{(2)}(\tau,\tau') = \Theta(\tau)\Theta(\tau')
\chi_{\alpha\beta\gamma}^{(2)}(\tau,\tau'). 
\end{equation}
By Fourier transforming Eq.~(\ref{eq:kk2}), we find that
\begin{align}
&\chi_{\alpha\beta\gamma}^{(2)}(\omega,\omega ') =
-\frac{\mathcal{P}}{3\pi^2} 
\int\limits_{-\infty}^{\infty}d\tilde{\omega}d\tilde{\omega}'
\frac{\chi_{\alpha\beta\gamma}^{(2)}(\tilde{\omega},\tilde{\omega}')}
{(\tilde{\omega}- \omega)(\tilde{\omega}' - \omega ')}
\nonumber\\ 
&-\frac{\mathcal{P}}{3\pi i}\int\limits_{-\infty}^{\infty}d\tilde{\omega}
\frac{\chi_{\alpha\beta\gamma}^{(2)}(\tilde{\omega},\omega ')}
{(\tilde{\omega}-\omega)}
-\frac{\mathcal{P}}{3\pi i}
\int\limits_{-\infty}^{\infty}d\tilde{\omega}'
\frac{\chi_{\alpha\beta\gamma}^{(2)}(\omega,\tilde{\omega}')}
{(\tilde{\omega}'-\omega')}.
\label{eq:kk3}
\end{align}
Splitting up Eq.~(\ref{eq:kk3}) into its real and imaginary parts gives
the nonlinear Kramers-Kronig relations of the form
\begin{multline}
\mathrm{Re}\left[\chi_{\alpha\beta\gamma}^{(2)}(\omega,\omega ')\right]
+ \frac{\mathcal{P}}{3\pi^2}
\int\limits_{-\infty}^{\infty}d\tilde{\omega}d\tilde{\omega}'
\frac{\mathrm{Re}\left[
\chi_{\alpha\beta\gamma}^{(2)}(\tilde{\omega},\tilde{\omega}')\right]}
{(\tilde{\omega} - \omega)(\tilde{\omega}' - \omega ')}
\\
= -\frac{\mathcal{P}}{3\pi}\int\limits_{-\infty}^{\infty}d\tilde{\omega}
\frac{\mathrm{Im}
\left[\chi_{\alpha\beta\gamma}^{(2)}(\tilde{\omega},\omega')\right]}
{\tilde{\omega} - \omega}
\\
- \frac{\mathcal{P}}{3\pi}\int\limits_{-\infty}^{\infty}d\tilde{\omega}'
\frac{\mathrm{Im}
\left[\chi_{\alpha\beta\gamma}^{(2)}(\omega,\tilde{\omega}')\right]}
{\tilde{\omega}'-\omega'},
\end{multline}
\begin{multline}
\mathrm{Im}\left[\chi_{\alpha\beta\gamma}^{(2)}(\omega,\omega ')\right]
+ \frac{\mathcal{P}}{3\pi^2}
\int\limits_{-\infty}^{\infty}d\tilde{\omega}d\tilde{\omega}'
\frac{\mathrm{Im}
\left[\chi_{\alpha\beta\gamma}^{(2)}(\tilde{\omega},\tilde{\omega}')\right]}
{(\tilde{\omega}-\omega)(\tilde{\omega}-\omega')}
\\
= \frac{\mathcal{P}}{3\pi}
\int\limits_{-\infty}^{\infty}d\tilde{\omega}
\frac{\mathrm{Re}
\left[\chi_{\alpha\beta\gamma}^{(2)}(\tilde{\omega},\omega')\right]}
{\tilde{\omega}-\omega}
\\
+\frac{\mathcal{P}}{3\pi} \int\limits_{-\infty}^{\infty}d\tilde{\omega}'
\frac{\mathrm{Re}
\left[\chi_{\alpha\beta\gamma}^{(2)}(\omega,\tilde{\omega}')\right]}
{\tilde{\omega}'-\omega'}.
\end{multline}
As in the case of linear response functions, all causal second order
response function must satisfy these relations.

The polarizability of an atom is a response function and hence must satisfy 
the relevant Kramers-Kronig relations. The full expression for the second 
order nonlinear polarizability in absorbing media is
\begin{align}
&\chi_{\alpha\beta\gamma}^{(2)}(\omega ,\omega ') =
\frac{1}{(i\hbar)^2}\frac{1}{\varepsilon_0}\sum_{ijk}\rho^{(0)}_{ii}
\nonumber \\ 
&\times\left[ \frac{d_{\alpha ,jk}d_{\beta ,ki}d_{\gamma ,ij}}
{(\omega '-\omega_{ik}-\delta\omega_{ik}-i\Gamma_{ik})
(\omega+\omega'-\omega_{ij}-\delta\omega_{ij}-i\Gamma_{ij})}\right.
\nonumber \\
&
\left. - \frac{d_{\alpha ,ij}d_{\beta ,ki}d_{\gamma ,jk}}
{(\omega'-\omega_{ik}-\delta\omega_{ik}-i\Gamma_{ik})
(\omega+\omega'-\omega_{jk}-\delta\omega_{jk}-i\Gamma_{jk})}\right.
\nonumber \\
&\left. -\frac{d_{\alpha ,ki}d_{\beta ,ij}d_{\gamma ,jk}}
{(\omega'-\omega_{ji}-\delta\omega_{ji}-i\Gamma_{ji})
(\omega+\omega'-\omega_{jk}-\delta\omega_{jk}-i\Gamma_{jk})}\right.
\nonumber \\
&
\left. + \frac{d_{\alpha ,jk}d_{\beta ,ij}d_{\gamma ,ki}}
{(\omega ' - \omega_{ji} - \delta\omega_{ji} - i\Gamma_{ji})
(\omega+\omega'-\omega_{ki}-\delta\omega_{ki}-i\Gamma_{ki})}\right]
\end{align}
which must obey
Eq.~(\ref{eq:kk3}).
Note that the response function is constructed from four
terms of the general form 
\begin{equation}
\label{eq:T}
T_{abd}(\omega,\omega ') = \frac{A}{(\omega -\omega_{ab}-i\Gamma_{ab})
(\omega + \omega ' -\omega_{ad}-i\Gamma_{ad})},
\end{equation}
where $A$ is a constant. Applying Eq.~(\ref{eq:kk3}) to (\ref{eq:T}) gives
\begin{widetext}
\begin{align}
T_{abd}(\omega,\omega ') &=  
-\frac{\mathcal{P}}{3\pi^{2}}\int^{\infty}_{-\infty}
d\tilde{\omega}d\tilde{\omega}'
\frac{T_{abd}(\tilde{\omega},\tilde{\omega}')}
{(\tilde{\omega}-\omega)(\tilde{\omega}'-\omega')} 
+ \frac{i\mathcal{P}}{3\pi}\int^{\infty}_{-\infty}d\tilde{\omega}
\frac{T_{abd}(\tilde{\omega},\omega ')}{(\tilde{\omega} - \omega)} 
+ \frac{i\mathcal{P}}{3\pi}\int^{\infty}_{-\infty}d\tilde{\omega}'
\frac{T_{abd}(\omega,\tilde{\omega}')}{(\tilde{\omega}'-\omega')},
\nonumber\\
T_{abd}(\omega,\omega ') &=  
-\frac{\mathcal{P}}{3\pi^{2}}\int^{\infty}_{-\infty}
d\tilde{\omega}d\tilde{\omega}'
\frac{A}{(\tilde{\omega} - \omega)(\tilde{\omega}' - \omega ')
(\tilde{\omega} - \omega_{ab}-i\Gamma_{ab})
(\tilde{\omega} + \tilde{\omega} ' -\omega_{ad}-i\Gamma_{ad})}
\nonumber\\
&\qquad +\frac{i\mathcal{P}}{3\pi}
\int^{\infty}_{-\infty}d\tilde{\omega}
\frac{A}{(\tilde{\omega} - \omega)(\tilde{\omega} - \omega_{ab}-i\Gamma_{ab})
(\tilde{\omega} + \omega ' -\omega_{ad}-i\Gamma_{ad})}\nonumber\\
&\qquad\quad +\frac{i\mathcal{P}}{3\pi}
\int^{\infty}_{-\infty}d\tilde{\omega}'
\frac{A}{(\tilde{\omega}' - \omega ')(\omega - \omega_{ab}-i\Gamma_{ab})
(\omega + \tilde{\omega}' -\omega_{ad}-i\Gamma_{ad})}.
\label{eq:Tint}
\end{align}
\end{widetext}
The integrals in (\ref{eq:Tint}) can be solved by residue calculus to
show that the rhs is indeed identically $T_{abd}(\omega,\omega')$.
Thus terms of this type, and hence the second order nonlinear
polarizability, obey the relevant Kramers-Kronig relations. Therefore,
the second order nonlinear polarizability is a causal response
function. 

\section{The Onsager Model}
\label{ons}

The Onsager model for local field corrections, as used in classical
nonlinear optics, is the classical variant of the real cavity model 
\cite{rcm,rcm2} and involves considering a point charge at the 
centre of a empty spherical cavity embedded within a dielectric medium. 
One then considers two cases: firstly, the field in the empty cavity
when the dielectric is subjected to an external electric field and,
secondly, the field in the cavity as a result of the polarization the
point charge induces in the surrounding dielectric. By considering the
Maxwell equations for the field on the boundary of the cavity, one can
derive relations between the applied and local fields electric and
polarization fields,
\begin{gather}
\mathbf{E}_\mathrm{loc} = \frac{3\varepsilon(\omega)}{2\varepsilon(\omega)+1}
\mathbf{E}\,,\\
\mathbf{P}_\mathrm{loc} = \frac{2}{3\varepsilon_0}\left[
\frac{\varepsilon(\omega)-1}{2\varepsilon(\omega)+1}\right]
\mathbf{P}.
\end{gather}
One should also note that since $\mathbf{P}_\mathrm{tot} = \mathbf{P} + \mathbf{P}_\mathrm{N}$,
the noise polarization field can be assumed to be corrected in the
same way as the reactive polarization field.

This method results in the same local field correction factors as the
real cavity model used for quantum systems. For a more detailed
description of the method readers are referred to Onsager's original
paper \cite{onsager}. A good summary of the model can also be found in
Ref.~\cite{onsager2}.

\end{document}